\documentclass[aps,pre,twocolumn,superscriptaddress,showpacs]{revtex4}
\usepackage{graphicx}
\usepackage{amssymb}
\begin{document}
\bibliographystyle{spm}

\title{Defining Contact at the Atomic Scale}
\author{Shengfeng Cheng}
\affiliation{Department of Physics and Astronomy, Johns Hopkins University,\\
3400 N. Charles Street, Baltimore, Maryland 21218, USA}
\author{Mark O. Robbins}
\affiliation{Department of Physics and Astronomy, Johns Hopkins University,\\
3400 N. Charles Street, Baltimore, Maryland 21218, USA}

\date{\today}

\begin{abstract}
Molecular dynamics simulations are used to study different definitions
of contact at the atomic scale.
The roles of temperature, adhesive interactions and atomic structure are studied for simple geometries.
An elastic, crystalline substrate contacts a rigid,
atomically flat surface or a spherical tip.
The rigid surface is formed from a commensurate or incommensurate
crystal or an amorphous solid.
Spherical tips are made by bending crystalline planes or removing
material outside a sphere.
In continuum theory the fraction of atomically flat surfaces that is in
contact rises sharply from zero to unity when a load is applied.
This simple behavior is surprisingly difficult to reproduce with atomic
scale definitions of contact.
Due to thermal fluctuations,
the number of atoms making contact at any instant rises linearly with load
over a wide range of loads.
Pressures comparable to the ideal hardness are needed to achieve full
contact at typical temperatures.
A simple harmonic mean-field theory provides a quantitative description of this
behavior and explains why the instantaneous forces
on atoms have a universal exponential form.
Contact areas are also obtained by counting the number of atoms with
a time-averaged repulsive force.
For adhesive interactions, the resulting area is nearly independent
of temperature and averaging interval, but usually rises from zero to unity
over a range of pressures that is comparable to the ideal hardness.
The only exception is the case of two identical commensurate surfaces.
For nonadhesive surfaces, the mean pressure is repulsive if there
is any contact during the averaging interval $\Delta t$.
The associated area is very sensitive to $\Delta t$ and
grows monotonically.
Similar complications are encountered in defining contact areas for spherical tips.
Even for the adhesive case, the area based on time-averaged forces can not be
described by continuum theory.
\end{abstract}

\pacs{46.55.+d,~62.20.Qp,~81.40.Pq}
% 46.55.+d   Tribology and mechanical contacts
% 62.20.Qp   Friction, tribology, and hardness
% 81.40.Pq   Friction, lubrication, and wear 

\maketitle

\section{Introduction}

The area of intimate contact between surfaces plays a
central role in continuum models of friction and adhesion.
Due to surface roughness,
this real area of contact $A_{\rm real}$ is typically much less than the
apparent area of the surfaces $A_0$.
Analytic and numerical work indicates that $A_{\rm real}$ rises
linearly with load for nonadhesive surfaces, whether they deform
elastically \cite{greenwood66,johnson85,persson01,hyun04} or
plastically \cite{bowden86,gao05,pei05}. 
This linear relation, and the assumption of a constant shear stress,
provides
one of the most common explanations of the linear relation
between friction and load in many experiments \cite{bowden86}.

The advent of nanotechnology and the ability to measure friction
in contacts with molecular dimensions has led to great interest
in extending continuum models to nanometer scales and
identifying its limits \cite{carpick97,carpick97b,enachescu98,carpick99,luan05,luan06,mo09,mo10,harrison99,robbins00c,knippenberg08,brukman08,pearson09,chandross08,yang08,yang08b}.
One fundamental question is what contact means at the molecular
scale.
The mere presence of an interaction between surfaces is not
enough, since attractive van der Waals interactions
extend to arbitrary scales.
One common approach has been to associate the onset of direct repulsion
between atoms with contact
\cite{luan05,luan06,mo09,mo10,knippenberg08,burnham91,luan09,cheng10pre}.
For most systems this coincides with common definitions of atomic
diameters \cite{footrepulse},
but the notion of contact is still complicated by
thermal fluctuations and the finite range of interactions.
These are not commonly included in continuum models and
recent work reveals that they greatly complicate the
extension of
continuum views of contact to atomic scales.

Early experiments found that friction measurements could be fit
by applying continuum theory to nanometer scale contacts and
assuming a constant shear stress at the interface
\cite{carpick97,carpick97b,enachescu98,carpick99,schwarz03}.
Simulations with tips of the same size showed that while friction forces
could be fit in this way, the fit parameters were often significantly
different than the independently determined material properties
\cite{luan05,luan06}.
The actual area of contact was often a factor of two larger than
the continuum prediction, the contact stiffness was up to
an order of magnitude smaller, and the friction force varied
by two orders of magnitude with changes in atomic geometry
by much less than an atomic diameter.

Knippenberg et al. \cite{knippenberg08} examined contact between
atoms in sliding contacts between nanometer scale tips and a substrate covered
with a surfactant.
They found that the area of contact was broadened due to the
compliance of the surfactant layer, but that most of the atoms contributed
little to the normal and lateral (friction) forces.
Roughly 90\% of the force was carried by only 10\% of the atoms,
calling into question the binary nature of contact.
Cheng et al. \cite{cheng10pre} considered the effect of
single monolayers of short chains.
They also found very large variations in the force on individual
atoms, both spatially and temporally.
One consequence was that measures that included
the magnitude of forces, such as moments of the pressure distribution,
gave very different contact radii than simply counting the atoms feeling
any nonzero repulsion.

Mo et al. have considered contact between
a bare diamond substrate and an amorphous surface that
is nominally spherical \cite{mo09,mo10}.
They found that the number of contacting atoms grew linearly
with the applied load.
They concluded that their surfaces were sufficiently rough
that the continuum results predicting area proportional
to load could be applied.
This idea was tested by comparing the distribution $P(f)$ of the
magnitude of local forces $f$ to Persson's continuum theory \cite{persson01}.

Yang et al. \cite{yang08,yang08b} have tested Persson's theory by simulating contact
between a flat substrate and a surface with roughness on all
length scales.
They found many more atoms with low forces than predicted.
The area of contact was not obtained by counting atoms, but
by fitting the distribution $P(f)$ at larger forces.
The result was then in reasonable agreement with Persson's
predictions.
Luan et al. \cite{luan09,luan05mrs,luan06a}
examined two dimensional models with much larger
linear dimensions.
They also found many atoms had low forces and that $P(f)$ decreased
monotonically with $f$.
The contact area obtained by counting atoms was extremely sensitive
to the detailed atomic structure.
One might not expect that continuum theory could capture these local
details, but it did correctly capture the long-range stress
correlations in the simulations \cite{luan09,campana08,hyun07,persson08c}. 

In this paper we explore the effect of thermal fluctuations on different
definitions of contact.
The simplest possible case, contact between atomically flat surfaces,
is considered first.
Even identical crystals with atoms directly over each other
exhibit unexpected behavior.
The number of atoms exhibiting repulsion at any instant rises linearly
with load, as might be expected for rough surfaces in continuum theory.
This linear scaling is observed for temperatures from $10^{-4}$ to 1/4 of the melting temperature, $T_m$. 
The contact area determined from time-averaged forces shows full contact
for this geometry, but not for incommensurate crystals with different lattice
constants or surfaces cut from amorphous solids.

A simple harmonic mean-field model for thermal fluctuations of 
individual surface atoms provides a quantitative
description of the above results.
It also explains several other new observations in our simulations.
These include a nearly universal exponential distribution
of the magnitude of instantaneous repulsive forces, and a corresponding
universal distribution of the fraction
of load carried by the atoms with the largest force.
There is also a direct correlation between the fraction of time in contact
and the time averaged force on an atom that is independent of surface structure.
The force required for atoms to remain in contact more than half
the time is surprisingly large at typical temperatures, with the corresponding pressure
comparable to the ideal hardness.

The prototypical case of contact between a spherical tip and flat substrate
is considered next.
Different atomic geometries of the tip, including commensurate, incommensurate
and amorphous are considered.
The same harmonic model used for flat surfaces describes
the time average force and fraction of time in contact for tip atoms.
For nonadhesive tips, the contact area
obtained by counting the number of atoms with a repulsive time-averaged force
depends on the observation time and grows monotonically.
Results for adhesive surfaces are less sensitive to time interval and
temperature, but differ substantially
from continuum theory.

The paper is organized as follows.
In Sec.~II we describe our simulation methods. 
Then MD results for the contact area between
flat surfaces are described in Sec.~III and a model
for the distribution of contact forces is developed
and tested.
Section IV examines contact between a sphere and flat
and
discussions and conclusions are presented in Sec.~V.

\section{Simulation Methods}

In continuum theory, contact between two rough elastic solids can be mapped to
that between a flat elastic substrate and a rigid rough upper
solid \cite{johnson85}.
We consider the latter case,
and take the upper solid
to be atomically flat or a spherical tip.
Previous studies at zero temperature indicate that the continuum mapping
remains approximately valid at atomic scales \cite{luan05},
but it does not take into account thermal effects.
At any finite temperature, annealed roughness from thermal fluctuations will be 
superimposed on top of any quenched structural roughness on the surfaces.
While the mapping may not accurately capture the quantitative effect of 
thermal fluctuations on both surfaces, using a rigid upper surface and a flat 
elastic substrate minimizes the parameter space to be explored.
Preliminary results for two elastic solids show the same effects
and trends are mentioned where relevant.

Atoms in the upper solid interact with those in the substrate via a 
truncated and shifted Lennard-Jones (LJ) potential \cite{allen87}:
\begin{equation}
\label{LJPotential}
V(r)=4\epsilon[(\sigma/r)^{12}-(\sigma/r)^{6}-(\sigma/r_c)^{12}+(\sigma/r_c)^{6}]~,
\end{equation}
where $r_c$ is the cutoff length, $\epsilon$ is the binding energy, 
and $\sigma$ is the atomic diameter.
We express all physical quantities in terms of LJ units based on 
$\epsilon$, $\sigma$, and the mass $m$ of the substrate atoms. 
For example, the unit of time is $\tau=\sqrt{m\sigma^2/\epsilon}$.
Where possible we plot dimensionless quantities in order to facilitate comparison with
experiments and theoretical models.

To model a nonadhesive contact, we take $r_c=2^{1/6}\sigma$ so that
the LJ potential is purely repulsive.
For adhesive contact, we use $r_c=2.2\sigma$ and reduce the binding energy to $0.5\epsilon$ to 
make adhesion between surfaces weaker than the internal cohesion in the substrate. 
This prevents the creation of cracks in the substrate under negative load (separating surfaces).

The substrate is a fcc crystal with a (001) surface. 
To make the substrate as elastic as possible, 
nearest neighbor atoms in the substrate interact through a harmonic potential: 
\begin{equation}
\label{HarmonicPotential}
V_{ij}(r)=\frac{1}{2}k(r-d)^2~,
\end{equation}
where $k$ is a spring constant and $d$ is the equilibrium spacing between nearest neighbors. 
We take $d=2^{1/6}\sigma$ to match the position of the minimum in the LJ potential in Eq.~\ref{LJPotential}.
The spring constant $k=57\epsilon/\sigma^2$ also matches the second derivative of the LJ potential
at its minimum.
Rather than listing the specific bonded pairs, spring forces are calculated
for all atoms that are within a distance of $1.3\sigma$.
This limits interactions to nearest neighbors, but does allow plastic
deformation to occur when forces are extremely high.
We limit our simulations to loads below this point.

The equilibrium density of the substrate at $k_{\rm B}T/\epsilon=0$
is $\rho=1.0m/\sigma^3$.
The initial state is a crystal of this density. Periodic boundary conditions with period
$\ell_x = \ell_y = 190.49\sigma$ are applied along the surface.
The crystal is $\ell_z=189.69\sigma$ high along the $z$ direction and the bottom layer is held fixed 
to mimic the support that balances the external load.
This depth is large enough to approximate contact of a sphere and a semi-infinite substrate \cite{luan06,adams06}.
The substrate has the same fcc structure and periodic boundary conditions 
in all simulations. As the temperature $T$ changes from
$10^{-4}\epsilon/k_{\rm B}$ (called the low $T$ case) 
to $T=0.175\epsilon/k_{\rm B}$ (called the high $T$ case), 
the height of the substrate shrinks by about $0.3\%$
because of a slight anharmonicity induced by changes in orientations of the springs. 
This slight breaking of cubic symmetry does not affect our results~\cite{cheng10pre}. 

We studied three atomic geometries for rigid flat surfaces.
Two contain a single (001) plane of a fcc crystal
with a lattice constant $d^{\prime}$. One is a commensurate surface with $d^{\prime}/d=1$.
This surface is in 
perfect alignment with the substrate, with its atoms
directly above the equilibrium positions of atoms in the top layer of the substrate. 
The second is an incommensurate surface with a lattice constant $d^{\prime}/d=1.12342$. 
The third surface has an amorphous structure. It is a thin sheet 
with thickness $\sim 0.8\sigma$ cut from an amorphous solid with density $1.0m/\sigma^3$.
The thickness was chosen to produce roughly the same number of surface atoms as 
the commensurate case.

The spherical tips have radius $R=100\sigma$ and four types of atomic
geometries studied previously \cite{luan05,luan06,cheng10pre}. 
The commensurate tip is made by bending a (001) plane of a fcc crystal with lattice constant $d'=d$
into a spherical shape.
The incommensurate tip is bent from a (111) plane of a fcc crystal with $d^{\prime}/d=1.12342$. The stepped 
tip is obtained by carving a spherical shell out of a commensurate 
fcc crystal with $d^{\prime}=d$. The detailed structure of the stepped tip depends on 
realization and in our case the bottom layer is a (001) face with $104$ atoms. 
The amorphous tip is a spherical shell carved 
out of the same amorphous solid from which the flat amorphous surface is cut. 
Tips used in AFM experiments are likely to be closest in structure to 
the amorphous or stepped tips.
The spherical tips and flat upper surfaces
all have a finite thickness. 
This is irrelevant for simulations of nonadhesive contacts.
For adhesive contacts, the finite thickness merely 
reduces the effective adhesion between two surfaces and 
does not affect the trends and conclusions presented below.

The simulations are performed using the Large-scale Atomic$/$Molecular Massively Parallel Simulator (LAMMPS) 
developed at Sandia National Laboratories. 
This classical MD code utilizes spatial decomposition to parallelize the computations.
Forces are calculated with the help of neighbor lists.
A velocity-Verlet algorithm with a time step $dt=0.005\tau$ is used to integrate the equations of motion.
The substrate is held at a fixed temperature $T$ using a Langevin thermostat. 
The Langevin damping rate $\Gamma$ is typically $0.1\tau^{-1}$, 
but results reported here are not sensitive to $\Gamma$ 
up to at least $\Gamma \sim 0.5\tau^{-1}$.

To illustrate the role of temperature, 
we report results for $T=10^{-4}\epsilon/k_{\rm B}$ 
and $0.175\epsilon/k_{\rm B}$. 
In some cases, we also report results for $T=0.07\epsilon/k_{\rm B}$. 
Note that the melting temperature of the substrates would be $T_m \sim 0.7\epsilon/k_{\rm B}$ 
if LJ interactions with the same length and stiffness were used. 
Thus the three temperatures above correspond to roughly $\frac{1}{7000}T_m$,
$\frac{1}{10}T_m$, and $\frac{1}{4}T_m$.
MD simulations ignore quantum effects that reduce thermal fluctuations.
Simulations below 5 to 10\% of $T_m$ are usually not representative of the
behavior of real materials, but may be useful for illustrating trends.

In most simulations, a constant normal load $L$ is applied to the top
surface and the system is allowed to equilibrate before data are collected.
The time for stress equilibration is a small multiple of the time for sound to
propagate across the system $\sim 20\tau$.
We equilibrated the system for at least $250 \tau$ after each change in
load, and data were typically averaged over a subsequent $500 \tau$.

For contact between nominally flat surfaces, the natural dimensionless
measure of load is $L/A_0 E^*$, where $A_0=\ell_x \ell_y$ is the nominal area
of the contacting surfaces, $L/A_0$ is the mean pressure
in the contact, and $E^*$ is the effective modulus.
The latter is related to Young's modulus $E$ and the Poisson ratio
$\nu$ by $E^*=E/(1-\nu^2)$ and $E^*=63 \epsilon/\sigma^3$ for
our substrate \cite{cheng10pre}.
The ratio $L/A_0 E^*$ is approximately equal to the average strain
along the $z$ direction far from the substrate.
For contact with a tip of radius $R$, the substrate dimensions
are effectively infinite and
the natural dimensionless measure of load is $L/R^2 E^*$.

\section{Contact between nominally flat surfaces}
\subsection{Nonadhesive Contact Area}

Molecular dynamics simulations provide complete information about forces and positions, 
but it is not obvious how this information should be used to identify the real area of contact.
For nonadhesive contact, it is natural to say an atom is in contact when it feels a force from
the opposing surface.
This can be generalized to adhesive contact by saying
atoms separated by less than some distance are in contact.
In the following we use the same distance as for nonadhesive contact,
so atoms are counted as contacting when the force between them is
repulsive.
This definition has also been used in discussing AFM experiments \cite{burnham91},
and previous simulations \cite{mo09,mo10,knippenberg08,brukman08,pearson09,cheng10pre,luan06a,harrison07}.
One can define an area of real contact, $A_c$,
by multiplying the number of contacting atoms $N_c$ by the area per atom $A_a$,
i.e., $A_c=N_c A_a$.
Thermal fluctuations introduce ambiguities in this definition because atoms fluctuate in
and out of contact.

We first consider the case of nominally flat non-adhesive surfaces.
 From the continuum perspective one would expect that these
surfaces should be in complete contact, $A_c=A_0$,
at any positive load.
However
surfaces constructed of discrete atoms can never be perfectly flat.
Even when all the rigid atoms are at the same height, as for commensurate and
incommensurate cases, the height at which substrate atoms feel a given
force varies with lateral position.
This corrugation depends on the relative size of substrate atoms and
the normal force, but the peak-to-peak height change is of order $\sigma/3$
for the cases considered here \cite{luan06,footBelidor}.
The change in height of atoms on the amorphous surface leads to additional
roughness of the same order.
Because the potential changes very rapidly with separation on
these scales, the atomic scale corrugation of any surface
can lead to large changes in contact properties even in the limit
of zero temperature considered in Refs. \cite{luan05,luan06}.

Thermal fluctuations lead to additional time varying or annealed roughness
on the surface.
An estimate of the magnitude of local fluctuations can be obtained
from the root mean squared (rms) normal displacement of a surface atom
if its eight neighbors are held fixed:
\begin{equation}
\delta z_{\rm rms} \approx \sqrt{k_{\rm B} T/k_{\rm eff}} \ \ ,
\label{zrms}
\end{equation}
where $k_{\rm eff}=2k$.
As $k_{\rm B} T/\epsilon$ increases from $10^{-4}$ to $0.175$,
this increases from about $10^{-3}\sigma$ to $0.04 \sigma$.
These values are similar to the measured
height fluctuations discussed further below.
One can estimate the period of normal vibrations $T_{vib}$
in the same way:
$T_{vib} \approx 2\pi \sqrt{m/k_{\rm eff}} \approx 0.6 \tau$.

While $\delta z_{\rm rms}$ is always smaller than the quenched
surface corrugations described above,
thermal fluctuations lead to time variation that significantly
complicates the definition of contact.
In particular, the number and identity of atoms that exert a force
on the opposing surface varies with time.
Some authors have associated $A_{real}$ with the mean number of
contacting atoms at any instant in time
\cite{mo09,mo10,knippenberg08}.
However, continuum theories generally consider time-averaged forces
and displacements.
 From this perspective it may be more natural to find the average
force on surface atoms over a long time interval
and identify all atoms with an average repulsion as in contact.
One can interpolate between these limiting cases by defining
$A_c(\Delta t)$,
the mean area associated with atoms that exert a repulsive force over
an averaging interval $\Delta t$, where $\Delta t$ varies from
a single time step to the length of the run \cite{cheng10pre}.
Note that this quantity does not measure any time evolution
or aging of the contact, but rather the mean number of contacting
atoms over a fixed
time interval averaged over all starting times in a steady
state simulation.

\begin{figure}[htb]
\centering
\includegraphics[width=3in]{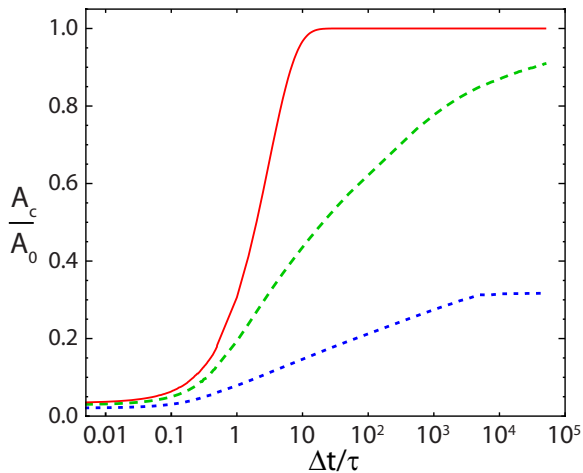}
\caption{(Color online) Fractional contact area $A_c/A_0$ vs the time interval $\Delta t$ on a log scale for three types of upper surfaces: 
commensurate (solid line); incommensurate (dashed line);
amorphous (dotted line). 
Here $k_{\rm B}T/\epsilon=0.175$
and $L/A_0 E^*= 5.5\times 10^{-4}$.
}
\label{ConAreaGro}
\end{figure}

Fig.~\ref{ConAreaGro} shows $A_c(\Delta t)/A_0$ for commensurate,
incommensurate, and amorphous surfaces at $k_{\rm B}T/\epsilon=0.175$
and $L/A_0 E^*= 5.5\times 10^{-4}$.
In all cases, only a few percent of the substrate atoms contact the rigid
surface at any instant of time.
This percentage is nearly constant for $\Delta t$ much shorter
than $T_{vib}$ because atoms have not had time to move \cite{cheng10pre}.
Over longer averaging intervals, $A_c$ rises and appears to saturate.
For the commensurate case, where all substrate atoms are directly under
rigid atoms, $A_c$ saturates at full contact in a time of order $10\tau$.
For the other surfaces there is a slow, roughly logarithmic, rise
over three decades in time before $A_c$ appears to saturate.
The final value from the average over the entire simulation
($5 \times 10^4 \tau$)
is over $90\%$ for the incommensurate surface,
and only $30\%$ for the amorphous case.

The difference in the asymptotic values of $A_c$ reflects the
atomic structure of the surfaces.
The commensurate case is the closest to ideally flat since all
substrate atoms are directly below rigid atoms and have the same
mean spacing.
As a result, all atoms
are likely to contact within
a time of order the vibrational time.
The above estimate of $T_{vib} \sim 0.6\tau$ was for short wavelength
modes.
While $A_c$ rises rapidly as the time interval becomes of this order,
complete saturation
does not occur until times of order the period of the slowest
long wavelength modes of the system $\sim \ell_z/c_s \approx 25 \tau$,
where $c_s$ is the speed of acoustic waves.

For the incommensurate case, each substrate atom has a slightly different
environment.
Some are directly below rigid wall atoms, and some are centered in between
wall atoms.
As noted above, this changes the height needed to produce a given force
by $\sim 0.3 \sigma$.
This is larger than the estimated thermal height
fluctuations, $\delta z_{\rm rms} \sim 0.04 \sigma$.
However, as $\Delta t$ increases,
more and more unlikely configurations will be sampled and the total
number of atoms that come into contact will grow.
In principle, all atoms should contact given an arbitrarily long time,
but over the times accessible to simulations the fractional
contact area grows roughly logarithmically and then appears
to saturate.
The situation for amorphous walls is similar, but the rougher
surfaces lead to a smaller long time value.

\begin{figure}[htb]
\centering
\includegraphics[width=3in]{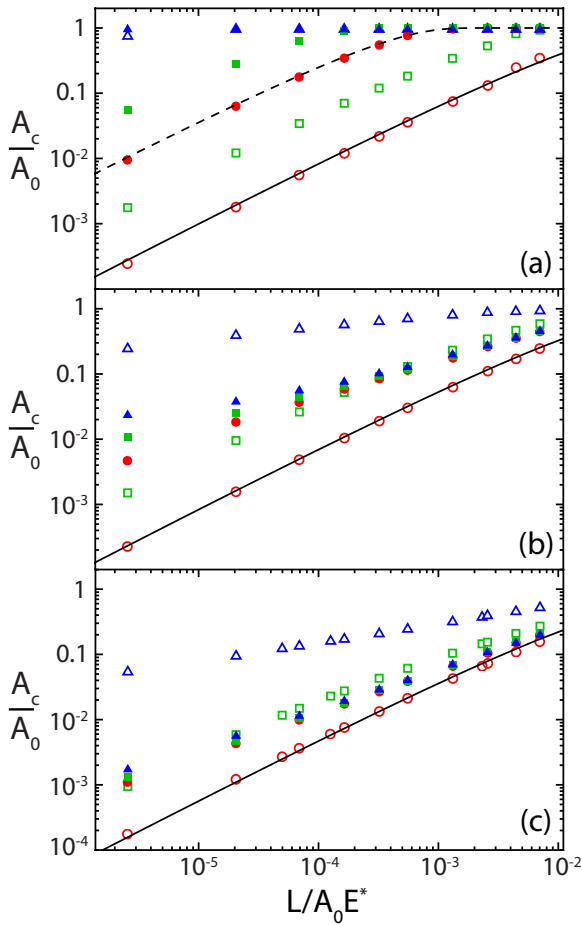}
\caption{(Color online) Contact area $A_c$ vs load $L$ for contact between a
nonadhesive elastic substrate and 
a rigid flat upper surface with different geometries: 
(a) commensurate; (b) incommensurate; (c) amorphous. 
Open and filled symbols are for $T=0.175\epsilon/k_{\rm B}$ and $10^{-4}\epsilon/k_{\rm B}$, respectively. 
The contact area is measured by counting the number of atoms in the top layer
of the substrate that interact with the 
opposite surface at any instant $\Delta t=0$ ($\bigcirc$)
or during time intervals $\Delta t=0.5\tau $ ($\Box$)
or $500 \tau$ ($\triangle$). 
The lines in the panel come from the model developed in Sec. III. C,
with solid lines and dashed lines for $k_B T/\epsilon =0.175$ and
$10^{-4}$, respectively.
}
\label{ConAreaPlaneNonAdh}
\end{figure}

Results for other temperatures and loads show qualitatively similar
behavior, but with different values of $A_c$ at early and late times.
To illustrate trends with $T$ and $L$, we will compare the mean results
for atoms contacting over three time intervals:
instantaneous $A_c(0)$,
a time interval $\Delta t=0.5\tau$ that is of order $T_{vib}$,
and a time comparable to the entire simulation $\Delta t = 500 \tau$
($\sim 1$ns).
Fig.~\ref{ConAreaPlaneNonAdh} shows results for commensurate,
incommensurate and amorphous walls
at two very different temperatures: 
$k_{\rm B}T/\epsilon =0.175$ (open symbols)
and $10^{-4}$ (closed symbols).
The data are plotted on a log-log scale to capture the wide range of values.
Note that averaging over even a single vibrational period of the atoms
has a dramatic effect on $A_c/A_0$, leading to an order of
magnitude increase in many cases.

At small loads, all curves are nearly linear, implying that
$A_c$ grows as a power of load.
The lines show fits to a model developed in Section \ref{sec:force} that
predicts a nearly linear relation between area and load.
Best fits to the data for $A_c(0)$ at high $T$ are consistent with this,
giving exponents a little greater than 0.9 over several decades.
Quenched geometrical disorder becomes more important at low temperatures,
leading to changes in the
apparent power law exponent for the amorphous and incommensurate cases.

For the commensurate surface, the area rises roughly linearly with load
for all temperatures and time intervals.
However the curves shift up with increasing $\Delta t$ since more
atoms have time to contact.
There is a corresponding downward shift in the load needed to reach full
contact.
Over a time of 500$\tau$ ($\sim 1$ns),
nearly all atoms contact even at the lowest load
and highest temperature.
This full contact is consistent with one's expectations for perfectly flat
surfaces in continuum theory.

Lowering the temperature increases the number of contacting atoms and
lowers the load needed to achieve full contact.
The rigid surface is held up by interactions with the highest
substrate atoms, because the repulsion changes rapidly with
separation.
The roughness from thermal fluctuations grows with temperature
as does the force from impacts at the thermal velocity.
These effects lead to a decrease in $A_c$ with increasing
$T$ at a fixed load that is captured by the model discussed below.

The incommensurate and amorphous surfaces show similar trends
at high temperatures, except that $A_c/A_0$ saturates below unity
on the time scale of our simulations.
The saturation value is always lower for the rougher amorphous
surface, but grows more rapidly with load.
At low $T$, the variation with time interval is much weaker,
with $A_c/A_0$ changing less than a factor of 2 with $\Delta t$ at all loads
for the amorphous case.
The reason is that the quenched structural variations due to the incommensurate
spacing or random atomic heights are large enough that some atoms are very
unlikely to be brought into contact by thermal fluctuations at low $T$.
Those substrate atoms that are closest to rigid atoms make contact quickly
and those farther away never contact.

It is interesting to compare the above results to a simple model
motivated by contact between
a solid wall and ideal gas with number density $n$.
When averaged over a long time, there is a uniform pressure
$p=n k_{\rm B}T$ on the wall from collisions with the gas.
 From the continuum perspective this implies that
the entire solid surface is in contact with the gas.
However at any instant, a very small fraction of solid atoms
feel a very large collision force.
The force is typically estimated from the change in momentum,
using a typical thermal velocity normal to the wall
$v_t \equiv \sqrt{k_{\rm B} T/m}$.
The momentum change in an elastic collision is $2mv_t$.
Taking the vibrational time as an estimate of the collision time,
the average force exerted on contacting atoms is
$\sim 2mv_t /T_{vib} = \sqrt{k_{\rm eff} k_{\rm B} T}/\pi$.
The fraction of area in contact will be the equilibrium pressure
times the area per atom $A_a$ divided by this force.
For typical values, the fractional contact area from this definition
is extremely small even though the entire surface contacts the
gas over long time intervals.

For nonadhesive solid-solid contacts with short range interactions
and high temperatures, the fraction of time in contact is
also small.
We can apply the same picture with the load
replacing the ideal gas pressure.
Then 
\begin{equation}
\frac{A_c}{A_0} = \frac{L A_a }{A_0} \frac{\pi}{\sqrt{k_{\rm eff} k_{\rm B}T}}=
c_A \frac{L}{A_0 E^*} \  \,
\label{idealgas}
\end{equation}
with the constant of proportionality 
\begin{equation}
c_A=\frac{\pi A_a E^*}{\sigma k_{\rm eff}}
\sqrt{\frac{\sigma^2 k_{\rm eff}}{k_{\rm B} T}}  \  \ .
\label{csubA}
\end{equation}
The first factor is of order unity since the same springs
determine both $E^*$ and $k_{\rm eff}$.
For $k_{\rm B}T/\epsilon =0.175$, $c_A \sim 56$ which is comparable to
the instantaneous values observed for all surfaces.
The slightly slower than linear increase in $A_c/A_0$ with load
is captured by the more detailed model developed in Sec. IIIC.

Decreasing $k_{\rm B}T/\epsilon$ from 0.175 to $10^{-4}$ should
increase $c_A$ by a factor of 42.
This is consistent with the data for commensurate surfaces.
The increase is slightly smaller for incommensurate surfaces and
even smaller for amorphous surfaces.
In contrast to the case of ideal gases, the quenched disorder on these
surfaces keeps some atoms from touching even at large times
and the number of these atoms grows as $T$ decreases.
As noted above, MD simulations ignore quantum effects that usually
become important below 5 to 10\% of the melting temperature.
Thus thermal fluctuations should be even smaller at this
low $T$, or equivalently it may be representative of the behavior
at higher temperatures when quantum effects are included.
Over the range of $T$ where MD simulations are accurate,
the behavior is close to that for the high temperature in
Fig. \ref{ConAreaPlaneNonAdh}.

To place the range of loads in perspective,
the mean pressures in the
contact, $L/A_0$, correspond to 0.5MPa (a few atmospheres) to 2GPa
for a typical metal with $E^* \sim 200$GPa.
The maximum normal pressure that can be supported before plastic deformation
is the hardness, $H$, and for bulk metals $H/E^*$ is typically
$10^{-4}$ to $10^{-3}$.
While significantly larger dimensionless hardnesses are observed in nanocrystals
and in amorphous materials, Fig. \ref{ConAreaPlaneNonAdh}
spans the range of dimensionless
loads that are likely to be found in real materials.

\subsection{Adhesive Contact Area}

The notion of contact is more complicated when long-range adhesive interactions
are present.
In principle, surfaces feel a van der Waals interaction at arbitrarily large
separations, although it will be arbitrarily weak.
Some threshold for interactions must be introduced to create a
sharp criterion for the onset of contact.
A common and reasonable choice is the onset of repulsive interactions \cite{burnham91}.
This corresponds to a separation distance at the minimum of the interatomic
potential, which is commonly used to define the atomic diameter.
It is also the same as the separation used to define
contact in the nonadhesive case for the LJ interactions used here.
Small changes in this criterion have little effect on the trends
described below.

\begin{figure}[htb]
\centering
\includegraphics[width=3in]{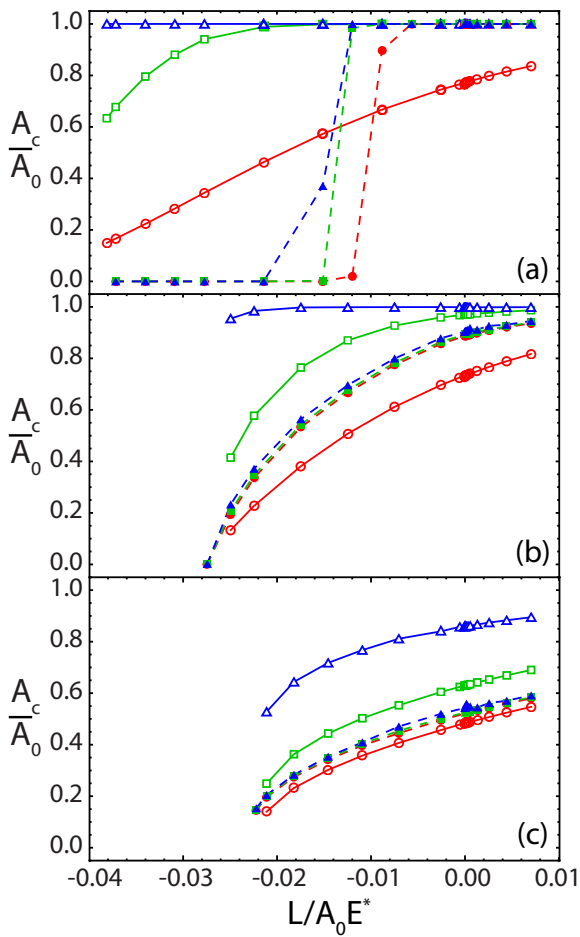}
\caption{(Color online) Contact area $A_c$ vs load $L$ for contacts between
an adhesive elastic substrate and 
a rigid flat upper surface with different geometries: 
(a) commensurate; (b) incommensurate; (c) amorphous. 
Open and filled symbols are for $k_{\rm B}T/\epsilon=0.175$ and $10^{-4}$,
respectively. 
The contact area is measured by counting the number of atoms in the top layer
of the substrate that feel a repulsion from the opposing
surface at any instant $\Delta t=0$ ($\bigcirc$)
or at any point during an interval of $\Delta t = 0.5\tau$ ($\Box$)
or $500 \tau$ ($\triangle$). 
}
\label{ConAreaPlaneAdh}
\end{figure}

Fig. \ref{ConAreaPlaneAdh} shows the contact area $A_c (\Delta t)$
for adhesive interactions between commensurate,
incommensurate and amorphous surfaces,
at different time intervals and temperatures.
A very strong adhesive potential corresponding to half the internal
cohesive potential is used to maximize the contrast with the nonadhesive
case.
This produces very large tensile strains of a few percent
that could produce yield if defects like dislocations were
present.
The range of interactions is extended to $r_c=2.2\sigma$, and
further extension had little effect on the results.

In all cases, adhesive interactions bind the surfaces in
a local free energy minimum.
The surfaces remain locked in this minimum until the load
exceeds a negative threshold $-L_p$, where
$L_p$ is often called the pulloff force in the context
of tip-substrate interactions.
The pulloff force is only unique in the limit of zero temperature,
where it represents the load where the energy minimum becomes
linearly unstable.
At any finite temperature, thermal activation will eventually
lead to pulloff at any constant negative load \cite{Israelachvili91}.
The pulloff force observed in simulations decreases with increasing
observation time and temperature, and increases with surface area.

As in simulations for tips \cite{luan06}, the pulloff force is largest
for the commensurate case where all substrate atoms are equally spaced
from rigid atoms and can exert the maximum adhesive force.
The pulloff force is lower for the incommensurate case where all
local environments are sampled, and even lower for the amorphous case
where additional roughness reduces the number of atoms that contribute
to the adhesion.
This effect is entirely geometric since the interaction potentials are
the same in all cases.

The behavior of $A_c$ at high temperatures is similar to that for
nonadhesive surfaces, although shifted to negative loads.
The fractional instantaneous contact area $A_c(0)$ rises roughly linearly
at low fractional contact areas
for the commensurate case, with increasing negative curvature
as area and load increase.
The curvature is more pronounced for the other surfaces, but note
that $A_c/A_0$ is bounded by unity and must saturate at large loads.
The linear scale of Fig. \ref{ConAreaPlaneAdh} accentuates the large values of
$A_c/A_0$, where there was also more curvature for the nonadhesive
case.
Thermal fluctuations in the pullof force prevent us from accessing
much smaller values of $A_c/A_0$ in constant load simulations,
but the asymptotic behavior is discussed further in Sec. IIIC.

For larger values of $\Delta t$, substrate atoms are counted as in
contact if they feel a repulsive force from a rigid atom at any
instant during the time interval.
This is consistent with the definition for nonadhesive surfaces and
leads to a monotonic rise in $A_c$ with $\Delta t$.
For the commensurate case, atoms are very likely to make contact
even within $0.5 \tau$ and full contact was obtained over the entire
simulation.
The incommensurate surface also reached full contact for most loads
at the longest time interval, while the greater roughness on the
amorphous surface led to saturation at lower $A_c/A_0$.

At low temperatures, thermal fluctuations are smaller and the
time interval is less important.
For the commensurate surface there is a sharp transition between
zero and full contact at $L/A_0 E^* \sim -0.01$.
The transition moves to negative loads as $\Delta t$ increases because
there is time for more distant atoms to contact.
As for nonadhesive surfaces, $\Delta t$ has much less effect
for incommensurate and amorphous surfaces, because
thermal fluctuations are smaller than the quenched
variation in atomic separation.

The above approach is only one way of generalizing the definition
of contact used for nonadhesive surfaces.
One could also say that atoms contact only when the time-averaged
force over $\Delta t$ is repulsive.
For nonadhesive surfaces the force is always repulsive so the time-averaged
force is repulsive if contact is made at any instant during $\delta t$.
For adhesive surfaces, intervals of repulsion can be countered by
attractive interludes.
 From the continuum perspective, it may be most natural to define contact
based on the time-averaged force.

\begin{figure}[htb]
\centering
\includegraphics[width=3in]{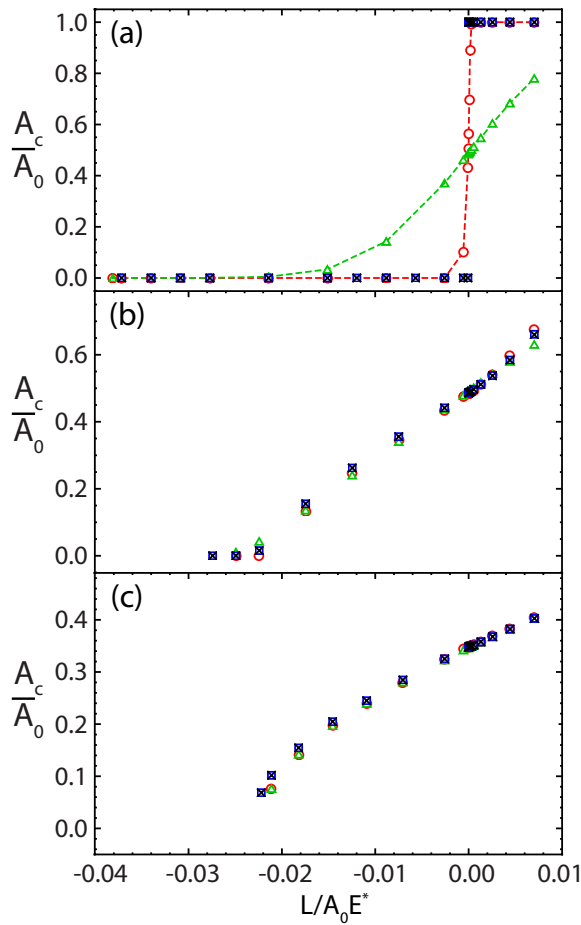}
\caption{(Color online) Contact area $A_c$ vs load $L$ for contacts between
an adhesive elastic substrate and 
a rigid flat upper surface with different geometries: 
(a) commensurate; (b) incommensurate; (c) amorphous. 
The contact area is determined from the mean number of atoms in the top layer
of the substrate that feel a time-averaged repulsive force from the opposing
surface over time intervals of $0.5\tau$ ($\bigcirc$ and $\Box$)
or $10^3 \tau $ ($\triangle$ and $\times$)
at
$k_{\rm B}T/\epsilon =0.175$ ($\bigcirc$ and $\triangle$) or
$10^{-4}$ ($\Box$ and $\times$).
}
\label{ConAreaPlaneAdhMean}
\end{figure}

Fig. \ref{ConAreaPlaneAdhMean} shows the fraction of area that feels
an average repulsion over time intervals of $0.5\tau$ and $500 \tau$.
The instantaneous results are the same as in Fig. \ref{ConAreaPlaneAdh}.
Averaging reduces the thermal fluctuations about the mean force.
For the amorphous and incommensurate surfaces, even averaging over 
$\Delta t=0.5\tau$ is sufficient to remove most of the fluctuations.
The quenched disorder then dominates the variation with load and
the results are nearly independent of both $T$ and $\Delta t$.

There is no quenched disorder for the commensurate case,
so all atoms have the same mean force.
For a sufficiently long time average, there is a sharp transition from
no contact to full contact at $L=0$.
Shorter time averages give a Gaussian distribution of mean
forces and $A_c/A_0$ rises like an error function.
Both the Gaussian and the rise in $A_c/A_0$ sharpen as $\Delta t$
grows.
Note that the width of this rise for $0.5\tau$ and a commensurate
surface, is smaller than the width of the rise for the other
surfaces and large $\Delta t$.
This is consistent with the lack of variation with $\Delta t$ for these
surfaces.

The above results indicate that defining contact area based on time
average forces provides a less ambiguous measure of contact
for adhesive systems.
Except in the extreme case of aligned, commensurate surfaces,
the results are insensitive to $\Delta t$.
In contrast, measures based on instantaneous contact show
an increasing drift with $\Delta t$ and greater variation
with $T$.
The lack of negative forces in the nonadhesive case means
that time average forces are sensitive to rare events.
This difficulty can be overcome by averaging the position rather
than the force.
Unlike the force, the position is not positive definite.
However, since the fraction of atoms in contact is almost always
much less than a half for the loads considered in the figures above,
criteria based on the mean position almost always give zero
contact area even though the local pressures are very high.

Note that the association of repulsion with contact is also somewhat
arbitrary.
One could argue that surfaces that are bound together in a free energy
minimum must be in contact.
In this case $A_c/A_0$ would be significant at any $L > -L_p$,
while Figures \ref{ConAreaPlaneAdh} and \ref{ConAreaPlaneAdhMean}
show zero contact area over much of this range.
This could be addressed by associating the distance for contact
with a larger spacing.
For example, the commensurate surface becomes mechanically unstable
when the atoms are separated by the distance $r \approx 1.244\sigma$
that maximizes the attractive LJ force.

\subsection{Force Distributions}
\label{sec:force}

The results described above show that thermal fluctuations lead to large
variations in the identity of atoms making contact and the force exerted
by these atoms.
Previous work has also shown that load may be very unevenly distributed between
atoms, with a very small fraction of the atoms carrying the bulk of the
load \cite{knippenberg08,cheng10pre}.
In this section we examine the distribution of forces in time and among
atoms.
The distributions are surprisingly universal and can be understood
from a simple model that treats each atom independently.

The distribution of forces at any instant
can be characterized by
the number of atoms with forces greater than some
threshold $f_t$, $N(f_t)$, and the total force from these
atoms $F(f_t)$.
Since we associate contact with repulsive (positive) forces, the
discussion will focus on $f_t \geq 0$, but it is readily
extended to negative values.
As $f_t$ decreases to zero, $N(f_t)$ and $F(f_t)$ approach the
total number of contacting atoms and the total repulsive force
at that instant, respectively.
The fraction of contacting atoms with $f>f_t$, $N(f_t)/N(0)$,
and the fraction of the repulsive force they carry, $F(f_t)/F(0)$,
both vanish as $f_t$ increases.

\begin{figure}[htb]
\centering
\includegraphics[width=2.5in]{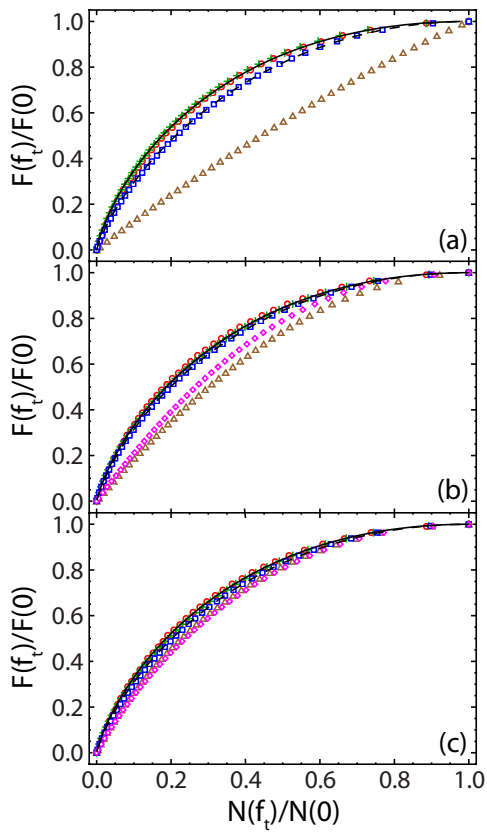}
\caption{(Color online) The fraction $F(f_t)/F(0)$ of load carried by atoms
with forces bigger than a threshold $f_t$ as a function
of the fraction of contacting atoms with these
large forces $N(f_t)/N(0)$ for (a) commensurate, (b) incommensurate
and (c) amorphous surfaces.
The threshold $f_t$ increases as one moves down and to the left
along each curve.
Circles and pluses are for nonadhesive surfaces at $k_{\rm B}T/\epsilon =0.175$
with loads of $L/A_0 E^* =2 \times 10^{-5}$ and $0.007$,
respectively.
Triangles are for $k_{\rm B} T/ \epsilon =10^{-4}$ and load $L/A_0 E^* =0.007$.
Squares and diamonds are for adhesive surfaces at
$k_{\rm B} T/\epsilon =0.175$ and $10^{-4}$ with
$L/A_0 E^* =0.007$.
The dashed lines in (a)-(c) are solutions of Eqns. \ref{eq:nft} and \ref{eq:fft} for
adhesive interactions and $k_{\rm B}T/\epsilon=0.175$.
The solid line in (a) is a similar solution for nonadhesive surfaces.
The solid lines in (b) and (c)
show the simple analytic approximation $y=x (1-\ln x )$,
which is very close to the full solution (the solid line in (a)).
}
\label{ForceStaSnapshot}
\end{figure}

As shown in Figure \ref{ForceStaSnapshot} we find that
the fraction of atoms carrying a given fraction of the total
repulsive force is surprisingly universal.
Results for all surfaces, with and without adhesion,
have nearly the same form at $k_{\rm B} T /\epsilon =0.175$.
Similar results are obtained at temperatures up to two orders
of magnitude lower.
Only at extremely low temperatures, such as
$10^{-4} \epsilon/k_{\rm B}$, does the amount of quenched disorder
become more important than thermal fluctuations.
Even in this limit the amorphous results remain
similar to the high temperature results.
Results for the perfectly ordered commensurate surface
approach a straight line, and the incommensurate results are
between the other two cases.

The results in Fig. \ref{ForceStaSnapshot} can be understood from a very
simple mean-field model.
The total potential for each substrate atom is approximated by an effective
harmonic spring $k_{\rm eff}$ binding it to the substrate plus the potential from
the rigid atoms.
Both are assumed to depend only on $z$, with repulsion from the rigid
atoms when $z>0$ and the equilibrium position relative to the substrate
at $z=-z_0$.
Then
\begin{equation}
U(z) = \frac{1}{2} k_{\rm eff} (z+z_0)^2+U_w(z) \  \ ,
\label{HarmPotential}
\end{equation}
where $U_w(z)$ is the wall potential.
For the commensurate case, all atoms see the same environment.
In the nonadhesive case, $U_w(z)$ is just the LJ interaction with
the rigid atom above, and in the adhesive case it can be calculated
by including the four next-nearest neighbors.
In principle, a more complicated interaction should
be used for amorphous and incommensurate simulations
because of variations in lateral position, but we will see that
the exact form of $U_w$ is not too important.

For a sufficiently large system we can replace the distribution over
atoms at a given
instant by the equilibrium Boltzmann distribution for a single atom.
The probability for an atom to be at height $z$ is then
\begin{equation}
p(z) = \frac{1}{Z}e^{-U(z)/k_{\rm B} T}\ \  ,
\end{equation}
with
\begin{equation}
Z = \int_{-\infty}^{+\infty} e^{-U(z)/k_{\rm B}T} {\rm d}z.
\end{equation}
Each threshold force $f_t$ corresponds to a threshold height $z_t$,
so one can write
\begin{equation}
N(f_t) = N_0 \int_{z_t}^\infty p(z) {\rm d}z,
\label{eq:nft}
\end{equation}
and 
\begin{equation}
F(f_t) = N_0 \int_{z_t}^\infty f(z) p(z) {\rm d}z,
\label{eq:fft}
\end{equation}
where $N_0$ is the total number of substrate surface atoms.
These integrals are readily performed numerically.
Results for all systems can be fit by varying $z_0$ at a
fixed $k_{\rm eff}=86 \epsilon/\sigma^2$.
This value of $k_{\rm eff}$ is slightly smaller
than the estimate of $2k$ obtained below Eq. \ref{zrms}
by considering fixed nearest-neighbors,
which reflects the extra compliance associated with the motion of these
neighbors.

Note that all surfaces follow nearly the same behavior at high
temperatures, and amorphous results are similar even at the
extremely low temperature of $10^{-4}\epsilon/k_{\rm B}$.
This surprisingly universal asymptotic behavior
can be understood from an even simpler approximation for 
small fractional contact area.
In this limit, $P(z)$ decays rapidly with $z$.
Most contacting atoms are near $z=0$ and
the integrals in Eqs. \ref{eq:nft} and \ref{eq:fft}
are dominated by the contribution near $z_t$.
For small $z$, the potential from the rigid wall
can be approximated by a quadratic function
$U_w(z) \approx k' z^2/2$ and $f(z) \approx k'z$.
In this limit,
$U_w$ is also much smaller than the potential from the substrate,
and the total potential
can be approximated by a linear function
\begin{equation}
U(z) \approx k_{\rm eff} z_0^2/2 + k_{\rm eff} z_0 z= k_{\rm eff}z_0^2/2 + z_0 k_{\rm eff}
f/k' \ \ .
\label{approx}
\end{equation}
Eq. \ref{eq:nft} can then be reduced to
$N(f_t) = c \int_{f_t}^\infty df \exp(-af) = (c/a) \exp(-af_t)$,
where $a= z_0 k_{\rm eff} /k' k_{\rm B} T$.
One also has
$F(f_t) = c \int_{f_t}^\infty df f \exp(-a f)
= -d N(f_t)/da = (1/a+ f_t) N(F_t)$.
Finally, the fractions become
$N(f_t)/N(0)= \exp(-a f_t)$ and
$F(f_t)/F(0)= (1+af_t) \exp(-a f_t)$.
The resulting solid curves in Figs. \ref{ForceStaSnapshot} (b) and (c)
have the form $y=x (1-\ln x)$ and are independent of $a$.
They differ only slightly from the
numerical integration of $F$ and $N$
at high $T$ shown by dashed lines in each panel and the solid line in (a).

\begin{figure}[htb]
\centering
\includegraphics[width=2.5in]{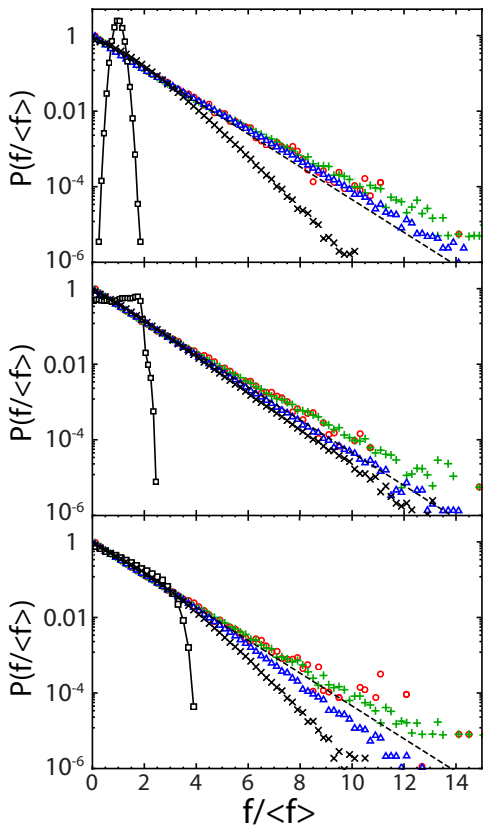}
\caption{(Color online) Probability density $P(f/\langle f \rangle)$ as
a function of the force on an atom normalized
by the average force $f/\langle f \rangle$.
Data are for three flat upper surfaces: (a) commensurate;
(b) incommensurate; (c) amorphous. 
For nonadhesive contacts, there are three data sets for
$k_{\rm B}T/\epsilon =0.175$
at $L/A_0 E^* = 2 \times 10^{-5}$ ($\bigcirc$), $5.5 \times 10^{-4}$ ($+$), 
and $0.007$ ($\triangle$), 
and one data set for $k_{\rm B}T/\epsilon=10^{-4}$ at
$L/A_0 E^* = 0.007$ ($\Box$).
For adhesive contacts, there is one data set for $k_{\rm B}T\epsilon=0.175$
and $L/A_0E^* = 0.007$ ($\times$).}
\label{exponentialforce}
\end{figure}

The key assumptions in the above approximation are that the
probability $P(z)$ decays exponentially for $z >0$ and that
the force $f=k'z$.
Fig. \ref{exponentialforce} shows that the distribution of
instantaneous local forces is exponential for a wide
range of circumstances.
The dependence of the distribution on $k'$, load and temperature
has been removed by normalizing by the mean force $\langle f \rangle$
obtained by averaging over
all atoms with an instantaneous repulsion.
The probability $P(f/\langle f\rangle)$ of a given multiple of this mean $f/\langle f\rangle$ is
then calculated by averaging over all atoms and times.

As expected from Fig. \ref{ForceStaSnapshot},
results for all high temperatures follow an exponential
form over several decades in $P$.
There are deviations at large $f$ because of anharmonicities
in the wall potential and variations in atomic separation,
but these are generally confined to regions of very low
probability that contribute little to average quantities.
The simple linear approximation to $U(z)$ provides a surprisingly
good description over the most important force range.
Similar results are obtained at $k_{\rm B}T/\epsilon =0.07$ or about 10\%
of the melting temperature.
Thus this exponential behavior should be present at room temperature
for nearly any solid.

Deviations from exponential behavior increase as $T$ decreases
further, and as the load increases.
In these limits,
the mean position $z_0$ moves towards and even past $z=0$.
The high load, high temperature behavior is most evident for
the adhesive cases in Fig. \ref{exponentialforce} where
adhesion leads to effective loads that are roughly 5 times higher
than the largest nonadhesive loads.
The effective mean normal pressure $L/A_0 \sim E^*/20$, would correspond
to $\sim 10$GPa for metals and the fractional contact area
is over 50\%
for all surfaces and definitions of contact (Fig. \ref{ConAreaPlaneAdh}).
Even here, the distribution is exponential for 2 decades or more.

The low temperature limit depends on geometry.
For the commensurate case, all atoms have the same potential and
probability.
For $k_{\rm B} T/\epsilon = 10^{-4}$ and $L/A_0 E^* =0.007$, $z_0$
is negative and atoms are
almost always in repulsive contact.
As a result, $f$ has a nearly Gaussian distribution about the mean value.
Atoms on the incommensurate surface sample different lateral
separations nearly uniformly.
The resulting force distribution is also nearly constant.
The random height distributions on the amorphous surface
give something closer to an exponential form, but cut off
at larger forces.
The disorder produces something like an effective temperature
in this case.

It is interesting to note that exponential distributions of forces
have also been observed in previous zero temperature studies
of surfaces that are even rougher than our amorphous surfaces.
Finite-element calculations of self-affine
fractal surfaces with roughness down to the
mesh size found a universal exponential decay \cite{hyun04},
although smoothing surfaces on small scales suppressed the distribution
at low and high pressures \cite{hyun07,campana07epl}.
Atomistic studies of fractal two dimensional surfaces with lengths
up to 8192 atoms also found an exponential decay \cite{luan09}.
Atomistic simulations of 3D rough surfaces \cite{mo09,mo10,yang08,yang08b}
were fit instead to a function predicted by Persson's
continuum theory \cite{persson01,hyun04}:
$P(x) = \pi x \exp(-\pi x^2/4)/2$.
However this analytic form is inconsistent with their observation
of a monotonic decrease in $P$ with $f/\langle f \rangle$.
Their plots of $P$ do not appear to be simply exponential either,
but were only plotted
for fractional contact areas of more than 10\%.
Their results are discussed further below.

Another recent study examined contact of amorphous surfaces at
a temperature of about 8\% of the melting temperature \cite{mo09,mo10}.
Their amorphous surface was rougher than ours, and they
found the force distribution was similar to Persson's
predicted form.
However, the distribution decayed less rapidly than Persson's
at large forces and only the forces from atoms that were
in contact more than 30\% of the time were included in
the distribution.
This choice was motivated by forcing the number of atoms
contributing to the distribution to equal the mean number in contact
at any instant \cite{mo09,mo10}.
As we show next, those atoms that spend least time in contact
support the smallest force and removing them from the
average will suppress $P$ at small forces.
It would be interesting to know whether the full force
distribution for all atoms in Mo et al.'s simulations had
an exponential form.

The quantities $f_c \equiv F(0)/N_0$ and $p_c \equiv N(0)/N_0$
represent the mean force
per atom and fraction of time in contact for a given separation
$z_0$ from the wall.
For the commensurate case, all atoms will have the same separation
and same average properties. 
As the load increases, $z_0$ will decrease, and
both $f_c$ and $p_c$ will increase.
Atoms on the amorphous and incommensurate surfaces have different
environments, and thus sample different $z_0$ at the same load.
If they act independently, then $z_0$ is the only relevant parameter
and results for $f_c$ vs. $p_c$ should fall on a universal curve
at each temperature.

\begin{figure}[htb]
\centering
\includegraphics[width=2in]{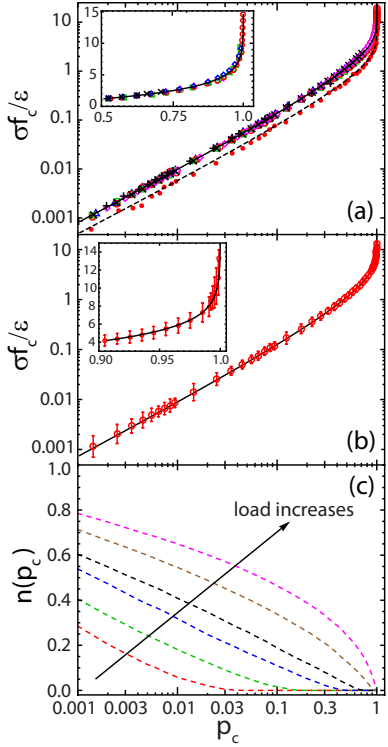}
\caption{(Color online)
(a) and (b) Time-average contact force on an atom $f_c=F(0)/N_0$ vs.
the fraction of time it is in contact $p_c= N(0)/N_0$.
(a) nonadhesive contacts of various upper surfaces: 
commensurate spherical tip ($\bigtriangleup$),
incommensurate spherical tip ($\Box$), amorphous sphetical tip ($\bigcirc$),
stepped spherical tip ($\bigtriangledown$),
commensurate flat surface ($+$), incommensurate flat surface ($\times$),
amorphous flat surface ($\Diamond$).
(b) adhesive contact of an amorphous flat surface.
The temperature is $T=0.175\epsilon / k_{\rm B}$
except for the indicated set of data at $T=0.07\epsilon / k_{\rm B}$.
The solid and dashed lines are fits using Eqs. \ref{eq:fft} and \ref{eq:nft}
with the wall potential and $k_{\rm eff}$ used in earlier figures.
(c) Fraction of atoms in contact more than a fraction $p_c$ of the time
as a function of $p_c$ for nonadhesive amorphous surfaces at loads
$L/A_0 E^* = 2.6\times 10^{-6}$, $2\times 10^{-5}$, $1.6\times 10^{-4}$, 
$5.5\times 10^{-4}$, $2.6\times 10^{-3}$, and 
$7\times 10^{-3}$ from left to right.
}
\label{ForceProb}
\end{figure}

Fig. \ref{ForceProb}(a) shows this collapse for nonadhesive surfaces.
For all systems, results for a range of loads are combined to sample
the full range of $p_c$.
The results are averaged over atoms within narrow ranges of $p_c$ and
the fluctuations are comparable to the symbol size.
The data are in excellent agreement with the simple harmonic mean-field
model, whose predictions for $k_{\rm B} T/\epsilon =0.175$
and 0.07 are shown by lines.
Note that the displacements associated with the largest forces are
quite large and are close to the threshold for plastic deformation.

Fig. \ref{ForceProb}(b) shows similar results for adhesive surfaces \cite{footrepel2}.
The agreement with simple theory is still quite good, but the fluctuations
between atoms (errorbars) are greater.
We find that the tails in the adhesive potential from nearby rigid atoms
lead to different shapes of $U_w$ for substrate atoms in different
environments, and there is also some coupling of lateral and normal
displacements.
While the distribution of force in Fig. \ref{ForceStaSnapshot}
is not affected by these changes in shape, the total forces and
probabilities are.
Note that the distributions for nonadhesive and adhesive contacts
in Fig. \ref{ForceProb} are fairly similar even though the LJ
interaction is half as strong in the adhesive case.
The reason is that interactions from multiple atoms are usually
important for the adhesive case and they roughly double the
effective strength of the potential.
This simple approximation is used in the fit line.

While the relation between $f_c$ and $p_c$ is independent of load,
the distribution of values for different atoms along a surface
changes with load.
For the commensurate case, all atoms have the same $p_c$ and $f_c$ in
long time averages.
For amorphous and incommensurate surfaces the quenched disorder leads
to a distribution of values corresponding to different $z_0$.
Fig. \ref{ForceProb}(c) illustrates how the range of $p_c$ changes
with load for the nonadhesive amorphous surface.
The fraction of contacting atoms $n(p_c)$ that are in contact more
than a fraction $p_c$ of the time is plotted against $p_c$ for
different loads.
As the load increases, the maximum values of $p_c$ (and thus $f_c$) increase.
In all cases, a large fraction of the contacting atoms spend a very small
fraction ($< 1$\%) of their time in contact.
This fraction of weak contacts drops from more than 90\% for the lowest load
to about 30\% for the highest load.
At larger $p_c$ there is roughly linear drop in $n(p_c)$ with the logarithm of
$p_c$.

\begin{figure}[htb]
\centering
\includegraphics[width=3in]{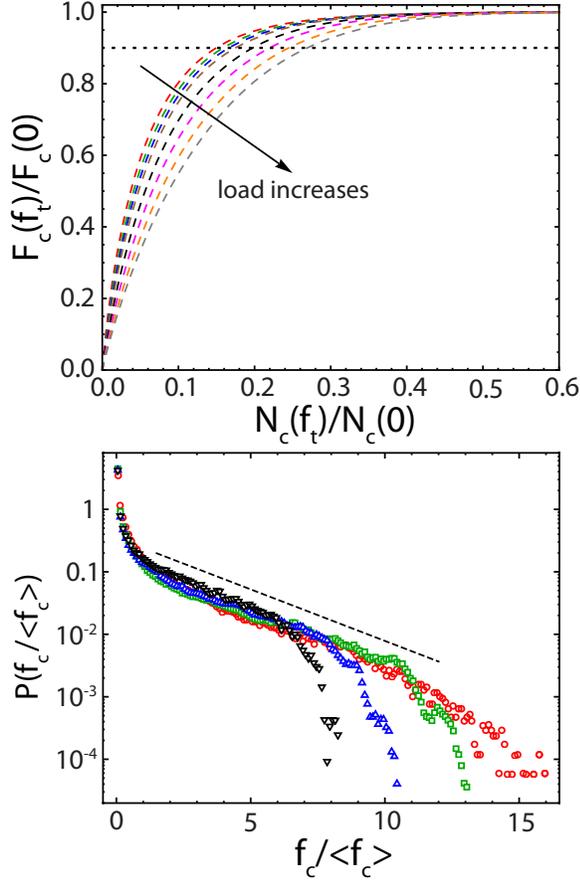}
\caption{(Color online) (a) Fraction of the repulsive force
$F_c(f_t)/ F_c(0)$ supported by the fraction $N_c(f_t)/N_c(0)$ of atoms with
time averaged force $f_c$ larger than a threshold $f_t$
($f_t$ decreases from left to right).
 From left to right the dimensionless loads are
$L/A_0 E^* = 6.9\times 10^{-5}$, 
$1.6\times 10^{-4}$, $3.2\times 10^{-4}$,
$5.5\times 10^{-4}$, $1.3\times 10^{-3}$, $2.6\times 10^{-3}$, 
$4.4\times 10^{-3}$ and $7\times 10^{-3}$.
(b) Probability $P(f_c/\langle f_c \rangle)$ of atoms having a time-averaged
force $f_c$ normalized by the mean force on all contacting atoms $\langle f_c \rangle$.
The loads are $L/A_0 E^* = 2\times 10^{-5}$ ($\bigcirc$), $5.5\times 10^{-4}$ ($\Box$), 
$2.6\times 10^{-3}$ ($\bigtriangleup$), and $7\times 10^{-3}$ ($\bigtriangledown$).
}
\label{timeave}
\end{figure}

The large number of weak bonds changes the distribution of forces from
the instantaneous results shown in Figs. \ref{ForceStaSnapshot}
and \ref{exponentialforce}.
By analogy with Fig. \ref{ForceStaSnapshot} we define $N_c(f_t)$ 
as the number of atoms with time-averaged force $f_c$ force greater than
a threshold $f_t$ and $F_c(f_t)$ as the force they carry.
Fig. \ref{timeave}(a) shows the fraction of load carried by the fraction
of atoms with the highest forces.
The fraction rises much more rapidly than for the instantaneous forces,
and depends upon load because the fraction of weak contacts drops
with load.
The fraction of atoms carrying $90\%$ of the the load rises from about
$15\%$ for the lowest load to $30\%$ for the highest load, as compared to
$60\%$ for the instantaneous force.
Knippenberg et al. have also found that a very small fraction of atoms
carries most of the load in their simulations \cite{knippenberg08}.

Fig. \ref{timeave}(b) shows the probability of an atom having a given
time average force $f_c$ as a function of $f_c$.
As in Fig. \ref{exponentialforce}, the force is normalized by the mean
force over all contacting atoms $\langle f_c \rangle $.
However, the curves are not independent of load.
In each case, there is a large peak at low forces corresponding
to the weak contacts, and these dominate the value of $\langle f \rangle$.
There is a nearly exponential distribution of forces among the
atoms that carry the majority of the load and Fig. \ref{timeave}(a)
can be fit reasonably well by assuming $P(f/\langle f \rangle)$ is
an exponential plus a delta function at zero force.
As noted above, Yang and Persson studied rougher surfaces and
found a $P(f/\langle f \rangle)$ that was monotonically decreasing.
Their data are also reasonably described by a strong peak at low
forces followed by an exponential region.

The above results have direct implications for the relation between
area and load found in earlier sections.
For the case of commensurate surfaces, all atoms have the same value of
$f_c$ and $p_c$.
The fraction of area in contact $A_c/A_0$ at a given load is just $p_c$
and the load is $L=f_c * N_0$.
The fit lines in Fig. \ref{ConAreaPlaneNonAdh}(a)
come from the same fit formula used in Fig. \ref{ForceStaSnapshot}
with no free parameters.
Making the same small area approximation as in Eq. \ref{approx},
one can derive an analytic
form for small $A_c/A_0$.
One obtains
\begin{equation}
\frac{N(0)}{N_0}=\frac{A_c}{A_0} = \frac {1}{\sqrt{2\pi}}\frac{1}{y} \exp (-y^2/2)
\label{simple1}
\end{equation}
and 
\begin{equation}
\frac{F(0)}{A_a N_0} = \frac{L}{A_0} = \frac{A_c}{A_0} \frac{k'\delta z_{\rm rms} }{ y A_a}
\label{simple2}
\end{equation}
with 
\begin{equation}
y=\sqrt{k_{\rm eff}z_0^2/k_{\rm B}T} = z_0/ \delta z_{\rm rms} \ \ .
\label{simple3}
\end{equation}
Since both $A_c$ and $L$ vary extremely rapidly with $y$, the value
of $y$ changes little over the relevant range of $A_c$.
This explains why $A_c$ rises only slightly less rapidly than linearly
with load in Fig. \ref{ConAreaPlaneNonAdh} at small $A_c/A_0$.
The main temperature dependence comes from $\delta z_{\rm rms}$ which
rises as $\sqrt{T}$, explaining the
scaling of $L$ with $\sqrt{T}$ in Fig. \ref{ConAreaPlaneNonAdh}(a).

For incommensurate and amorphous surfaces the situation is complicated,
because the distribution of $z_0$ is not known.
For sufficiently high temperatures, thermal fluctuations are more
important than geometrical fluctuations and $A_c$ rises roughly
linearly with load at small loads.
However, some fraction of the substrate atoms are so far from
wall atoms that they never contact during the simulation.
This can be incorporated by reducing $N_0$ to the number of substrate
atoms that can contact.
The fits in Fig. \ref{ConAreaPlaneNonAdh} reduce the number of atoms
to 84\% and 57\% of $N_0$ for incommensurate and amorphous surfaces,
respectively.
These fractions are consistent with the number of contacting atoms
in the limits of large load and time interval and provide an excellent
fit.
The corresponding predictions for $k_{\rm B}T/\epsilon = 10^{-4}$,
would be substantially above the actual data.
The reason is that many fewer atoms are close enough to contact,
particularly at low load.

\section{Contact between a spherical tip and a flat substrate}

The case of a spherical tip can also be analyzed with the theory
developed in the previous section.
The main difference is that the curvature of the tip leads to
additional variations in the separation $z_0$ between substrate
and wall atoms.
These variations increase in magnitude and importance as the
radius of the tip decreases.
We will consider a relatively small radius, $R =100\sigma$.
This is comparable to
the radius of AFM tips, leading to contacts with relatively small
numbers of atoms and rapid changes in surface separation.

Fig. \ref{ForceProb}(a) includes results for the mean force
and fraction of time in contact from simulations with nonadhesive
spherical tips.
Results for all geometries collapse on the same universal curve
obtained for flat surfaces.
A similar collapse with data in
Fig. \ref{ForceProb}(b) is found for adhesive tips.
Tests of the distribution of instantaneous forces confirm that
the same independent atom model describes substrate atoms
under all tips, although the small number of atoms in the
contacts means that results from several longer simulations are
required to get similar statistical accuracy.
The main new features of the results presented below
come from changes in $z_0$ due to surface curvature.

The continuum limit of nonadhesive contact by a spherical tip
is described by
Hertz theory \cite{johnson85}.
It predicts that the region of
contact is a circle whose radius $a$ scales as
\begin{equation}
\label{HertzRadius}
\frac{a}{R}=\left(\frac{3L}{4E^{*}R^2}\right)^{1/3} \ \ .
\end{equation}
The contact area is then $\pi a^2$ and is thus proportional to $L^{2/3}$,
rather than rising linearly with load.
The pressure in the contact depends on the radius $r$ from the center
and is given by:
\begin{equation}
\label{HertzPress}
p(r)/E^*=\frac{2}{\pi} \left( 3L/4E^* R^2 \right)^{1/3} \sqrt{1-r^2/a^2}  .
\end{equation}
Note that this dimensionless pressure and the radius
are proportional to the same dimensionless measure
of load $(L/E^* R^2)^{1/3}$.
The characteristic force per atom $f_a = p(0) A_a$
also grows as the cube root of load, while it is linear
in load for flat surfaces.
This makes it hard for tip simulations to span a wide
range of forces, and nanometer scale tips tend
to produce forces at the large end of Fig. \ref{ForceProb}.

Luan and Robbins considered the same spherical tips used here
in the zero temperature limit \cite{luan05,luan06}.
They found systematic deviations between the time and
angle averaged
pressure $p(r)$ and the Hertz prediction (Eq. \ref{HertzPress}).
The finite range of interactions and the atomic scale roughness
smeared the pressure over a larger area, and $p$ remained
nonzero to radii that were twice the predicted $a$ at
low loads.
Deviations were much larger for stepped tips, where the
area decreased in discrete jumps as new terraces made
contact.

\begin{figure}[htb]
\centering
\includegraphics[width=2.5in]{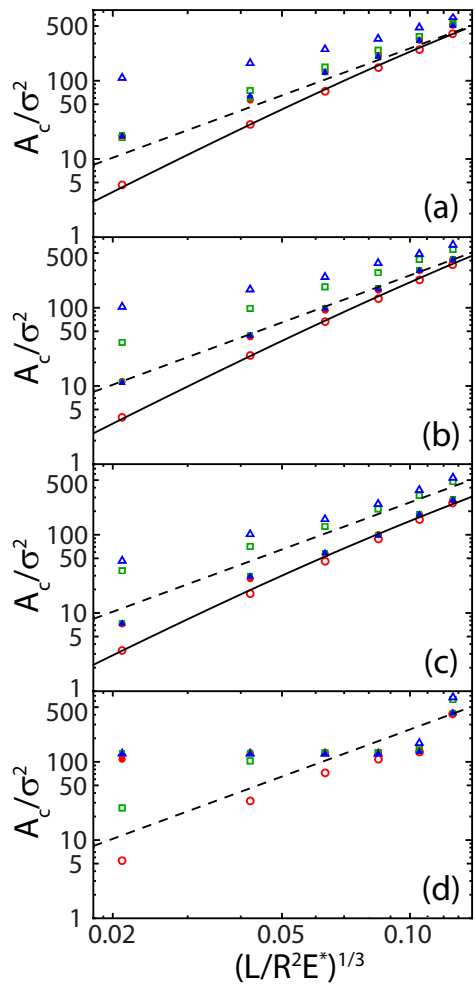}
\caption{(Color online) Contact area $A_c$ vs load $L$ for a spherical tip with different geometries: 
(a) commensurate; (b) incommensurate; (c) amorphous; (d) stepped. 
Open and filled symbols are for $T=0.175\epsilon/k_{\rm B}$ and $10^{-4}\epsilon/k_{\rm B}$, respectively. 
The contact area is measured by counting the number of atoms in the top layer
of the substrate that interact with the 
opposite surface at any instant $\Delta t=0$ ($\bigcirc$)
or during time intervals $\Delta t=0.5\tau $ ($\Box$)
or $500 \tau$ ($\triangle$).
The dashed lines represent the Hertz prediction and
are the same in all panels.
Solid lines represent the fits for each tip from the simple harmonic
mean-field theory
developed in Sec.~\ref{sec:force} with $k_{\rm B}T/\epsilon=0.175$ and
$N_0$ set equal to the number that contact at $k_{\rm B} T/\epsilon =10^{-4}$.
}
\label{ConAreaTipNonAdh}
\end{figure}

Figure \ref{ConAreaTipNonAdh} shows the
load dependence of the area $A$ obtained by counting
contacting atoms over different time intervals
as a function of $(L/E^* R^2)^{1/3}$.
The Hertz prediction is also shown by a dashed line.
This prediction is the same for all tips since their surfaces
differ by less than a molecular diameter.
However, they show rather different behavior, particularly in
the case of the stepped tip.

As for the flat surface results in the previous
section, the low temperature behavior is relatively
insensitive to the averaging time and the results are most strongly
influenced by atomic geometry.
The contact area for the stepped tip is equal to that of the
bottom terrace at low loads and jumps up at high loads when the
second terrace contacts.
The other tips all show the same scaling as the Hertz prediction
at low loads, but with different prefactors.
As shown by Luan and Robbins, the contact is spread over a larger
radius than predicted by Hertz theory \cite{luan05,luan06}.
All atoms of the commensurate surface that lie within this region
make contact, leading to an area about twice as large as the Hertz
prediction in Fig. \ref{ConAreaTipNonAdh} (a).
The quenched variation in substrate-tip separations for incommensurate
and amorphous surfaces reduces the fraction of atoms that contact.
This almost exactly compensates for the increased radius of contact 
for the incommensurate surface, while the number of contacting
atoms is roughly half the Hertz prediction for the amorphous surface.
The ratio of MD results to Hertz predictions depends
sensitively on the ratio of lattice constants
in the incommensurate case \cite{luan09}, and to the density in the amorphous
case, but is typically within a factor of two.

The time interval is much more important for high temperatures.
The area corresponding to atoms in contact at any instant is substantially
reduced from Hertz theory in all cases.
The rise with load is also more rapid than the Hertz prediction,
showing something like the more linear area-load relation found
for flat surfaces.
Note that the range of dimensionless pressures is much higher
for tips because of the small radius.
The Hertz prediction for our geometry and loads gives
$p/E^*$ between 0.01 to 0.07, while
for flat surfaces
$p/E^* \sim L/A_0E^*$ ranged from less than
$10^{-5}$ to 0.01 in Fig. \ref{ConAreaPlaneNonAdh}.

The solid lines in Fig.~\ref{ConAreaTipNonAdh}(a)-\ref{ConAreaTipNonAdh}(c) show the prediction of the harmonic mean-field theory
for the commensurate, incommensurate, and amorphous tips, 
taking the corresponding low temperature area as $A_0$.
The results are in excellent agreement with the simulations
and imply that the contact area of tips should scale linearly with load
at low $p/E^*$.
It is not possible to test this scaling to lower loads in our simulations
since there are only a handful of atoms touching at the two lowest loads.
Accessing the lowest $p/E^*$ studied for flat surfaces would require
increasing $R$ into the micrometer range.
This size scale is relevant for nanoindenters and is consistent with
common estimates of the size of surface asperities in macroscopic
contacts \cite{bowden86,dieterich96}.

As the averaging time interval $\Delta t$ grows,
the contact area rises above the Hertz
prediction.
The increase is more than an order of magnitude at low loads.
The results also become much less sensitive to tip geometry,
because thermal fluctuations overcome the quenched variation
in height.
The magnitude of the increase in area can be estimated from the
tip geometry and our estimate of height fluctuations
in Eq. \ref{zrms}.
Near the edge of a contact of radius $a$, the height varies with
change in radius $dr$ as $h(r) \approx h(a)+a dr/ R$.
Thus a fluctuation in height by $dz$ can allow contact
out to a radius that is larger by $dr \approx R dz/a$
and an area that is larger by $dA=2\pi a dr = 2\pi R dz$.
If we assume that $dz$ is about three times the standard
deviation $\delta z_{\rm rms}$ over the course of the simulation,
then $dA \sim 70 \sigma^2$ for $R=100\sigma$ and
$\delta z =0.04 \sigma$.
The observed changes are comparable to these simple estimates.
The fractional change in area scales as $\delta z_{\rm rms} /R$ and
would become smaller for micrometer and
larger tips.

The time averaged pressure is small in the region where contact
is only made possible by thermal fluctuations.
One consequence is that measures of the contact radius based
on the second moment of the pressure distribution remain
closer to the value predicted by Hertz and measured at low temperature
\cite{cheng10pre}.
Another is that a small fraction of the atoms carry a very
large fraction of the load.
This observation is similar to the result for time averaged
forces between flat surfaces at low temperatures, but is due to the
large ring at the perimeter of the contact that is within $\delta z$
of the mean substrate height.

\begin{figure}[htb]
\centering
\includegraphics[width=2.5in]{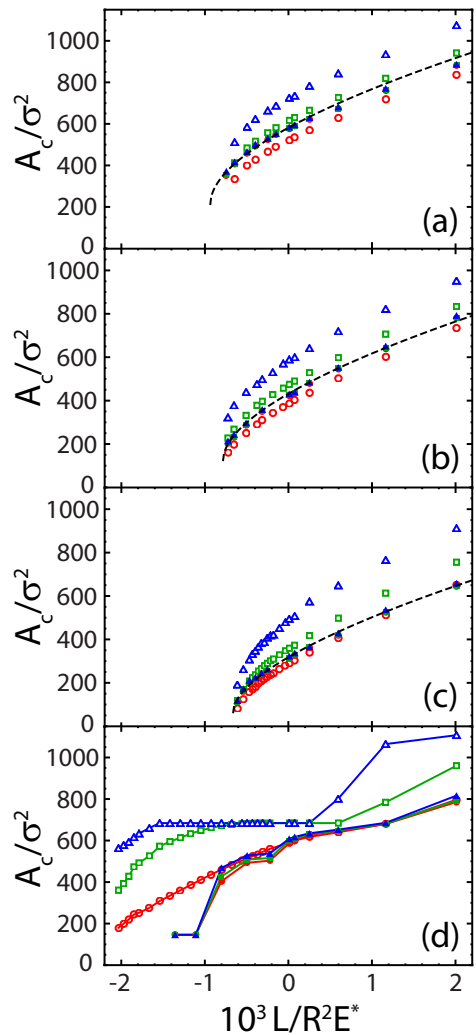}
\caption{(Color online) Contact area $A_c$ vs load $L$ for a spherical tip with different geometries: 
(a) commensurate, (b) incommensurate, (c) amorphous, and (d) stepped. 
Open and filled symbols are for $T=0.175\epsilon/k_{\rm B}$ and $10^{-4}\epsilon/k_{\rm B}$, respectively. 
The contact area is measured by counting the number of atoms in the top layer
of the substrate that ever repel the 
opposite surface at any instant $\Delta t=0$ ($\bigcirc$)
or during time intervals $\Delta t=0.5\tau $ ($\Box$)
or $500 \tau$ ($\triangle$).
Also shown is a fit based on the M-D theory to the data for $T=10^{-4}\epsilon/k_{\rm B}$.}
\label{ConAreaTipAdh}
\end{figure}

Figure \ref{ConAreaTipAdh} shows corresponding results for adhesive tips.
Results for all but the stepped tip look very similar to the flat amorphous
case.
The area rises sublinearly near the pulloff force, and the curves become
more linear at large loads.
The time interval has little effect at low temperatures because height
fluctuations are smaller than geometrical features.
The instantaneous contact area is smaller at high temperatures, but
$A_c$ grows with $\Delta t$ as thermal fluctuations bring
more atoms into contact.

What is intriguing is that while the results for adhesive tips look similar
to those for flat amorphous solids, they are also qualitatively similar
to continuum predictions for sphere-on-flat contact.
The form of the continuum theory depends on the
work of adhesion per unit area $w$ and the range of interactions $h_0$.
The key dimensionless parameter scales as \cite{maugis88,maugis90,tabor76}
\begin{equation}
\lambda \equiv \left( \frac{9Rw^2}{2\pi E ^{*2}h_0^3} \right)^{1/3}  \ \ .
\label{MDPara}
\end{equation}
The limit of strong, short range interactions ($\lambda > 5$) is described by
Johnson-Kendall-Roberts (JKR) theory,
while the opposite limit ($\lambda < 0.1$) is
described by Derjaguin, Muller and Toporov (DMT) theory.
For typical scanning probe microscope tips and our
simulations, $\lambda \sim 0.1$ to 1 lies between
JKR and DMT limits.
Maugis-Dugdale (M-D) theory provides a description of this
intermediate region, and approximate analytic expressions
for this theory have been developed to simplify fits to
M-D theory \cite{carpick99,schwarz03}.

\begin{table*}
\caption{Parameters from fits of contact area in Fig.~\ref{ConAreaTipAdh} 
to M-D theory.
Independently determined values for $w \sigma^2/\epsilon$
at zero temperature are
1.05, 0.45 and 0.23 for commensurate, incommensurate and amorphous cases.}
\tabcolsep0.1in
\begin{tabular}{cccccc}
\hline\hline
tip & parameter & \multicolumn{3}{c}{$T=0.175\epsilon/k_{\rm B}$} & $T=10^{-4}\epsilon/k_{\rm B}$ \\
 & & $0.005\tau$ & $0.5\tau$ & $500\tau$ & \\
\hline
commensurate & $L_p (\epsilon/\sigma) $ & 592 & 573 & 558 & 589 \\
 & $w (\epsilon/\sigma^2)$ & 1.11 & 1.21 & 1.38 & 1.19 \\
incommensurate & $L_p (\epsilon/\sigma) $ & 480 & 477 & 483 & 493 \\
 & $w (\epsilon/\sigma^2)$ & 0.84 & 0.93 & 1.08 & 0.90 \\
amorphous & $L_p (\epsilon/\sigma) $ & 391 & 331 & 332 & 417 \\
 & $w (\epsilon/\sigma^2)$ & 0.65 & 0.64 & 0.79 & 0.70 \\
\hline\hline
\end{tabular}
\end{table*}

Except for the stepped tip, the data shown in Fig.~\ref{ConAreaTipAdh}
can be fit reasonably well to M-D theory if the pulloff force $L_p$
and work of adhesion $w$ are treated as fit parameters. 
Fit parameters are summarized in the Table I.
While the fits look reasonable, the surfaces often separate at forces
that are significantly less negative than the fit $L_p$.
This is particularly evident for Fig~\ref{ConAreaTipAdh}, where the system
was not stable at any value lower than the final data point. 
In addition, the fit values of $w$ are all larger than the
independent measures of $w$ obtained
previously for the same surfaces at zero temperature (caption) \cite{luan06}.
Presumably this increase in the fit $w$ is needed to compensate for the
increase in contact area due to the atomistic effects discussed
above.
The fit parameters are generally within a factor of two of
independently measured quantities and may thus be useful
in extracting approximate values from experiments.
However, our results reinforce the conclusion reached previously
\cite{luan05,luan06,mo09}
that the success of fits to M-D theory does not represent a
quantitative success of continuum theory, which can not be expected
to hold at atomic scales.

\begin{figure}[htb]
\centering
\includegraphics[width=3in]{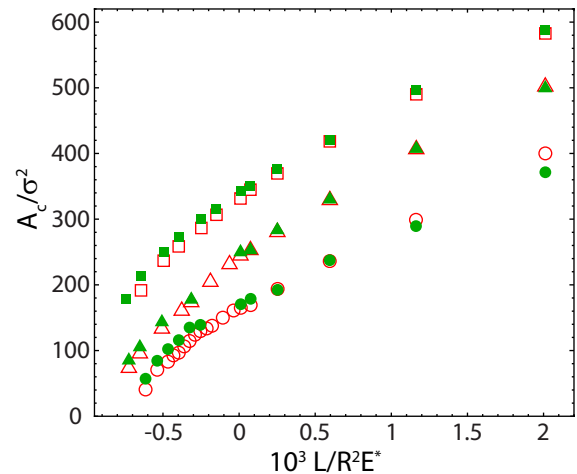}
\caption{(Color online) Contact area $A_c$ based on average force (contact means
average force being repulsive) vs load $L$ for different tip geometries:
commensurate ($\Box$), incommensurate ($\triangle$), and amorphous ($\bigcirc$).
Open (filled) symbols are for $T=0.175\epsilon/k_{\rm B}$ ($10^{-4}\epsilon/k_{\rm B}$).
}
\label{ConAreaTipAdhAve}
\end{figure}

The contact areas in Fig.~\ref{ConAreaTipAdh} were obtained
by counting all atoms that ever felt a repulsion during
the given $\Delta t$.
Fig.~\ref{ConAreaTipAdhAve} shows the area obtained
by counting only atoms
that have a net repulsive force after averaging over $\Delta t$.
As in the case of contact between two flat surfaces, this definition
of $A_c$ is insensitive to both averaging time interval and temperature.
It is also harder to describe with continuum theory.
The area grows smoothly with load, but could not be fit to M-D theory.
The reason is that the power law describing the rate of increase
in area is significantly different from the value
of 2/3 that enters continuum theories.
For flat incommensurate and amorphous surfaces the area rose smoothly
rather than transitioning rapidly to full contact.
A similar effect seems to modify the rise in area with load for the
case of spherical tips.
As for flat surfaces the suppression is larger for amorphous surfaces
than incommensurate surfaces.

\section{Discussions and Conclusions}

The results presented in this paper reveal the difficulties in extending
the notion of contact to molecular scales, even for the simplest geometries.
In the case of parallel, atomically flat surfaces, continuum theory would
predict a sharp transition from no contact to full contact.
The transition would occur at zero load for nonadhesive surfaces,
and also for adhesive surfaces if contact is identified with repulsion.
This sharp transition is only seen in atomic simulations for
the highly unlikely case of two identical, commensurate surfaces
with atoms aligned on top of each other.
Even there, it is only found near zero temperature or using time averaged
forces with adhesive interactions.
In all other cases, the contact area rises very gradually.
Indeed, pressures comparable to the ideal
hardness are required to reach full contact.

Measures of contact area based on instantaneous forces between atoms
were shown to be extremely sensitive to thermal roughness.
One common approach is to identify the contact area $A_c$ with the mean
number of atoms feeling a repulsive force times the area per atom
\cite{mo09,mo10,knippenberg08,luan09,luan06a}.
For both flat surfaces and spherical tips,
thermal fluctuations lead to a linear rise in $A_c$
with load when the mean pressure in the contact is small.
For flat commensurate surfaces, linear scaling extends over
the entire temperature range studied,
corresponding to $T$ from 0.01\% to 25\% of the melting temperature.
For incommensurate and amorphous surfaces, the linear behavior
is suppressed at low temperatures, as the variation in local
geometry around surface atoms becomes more important than
thermal fluctuations.
At $k_B T/\epsilon = 10^{-4}$, $A_c$ rises sublinearly,
but still remains below full contact until very high loads.

A simple mean-field model was developed that quantitatively describes
changes in $A_c$.
Each substrate atom is assumed to be bound to the substrate
by a harmonic spring $k_{\rm eff}$ that represents
the net effect of bonds to neighboring substrate atoms.
The atom also feels a force from the opposing wall $U_w(z)$
that depends on its height $z$.
The probability of an atom having a given height and force
can then be calculated from the Boltzmann equation.
Over a wide range of parameters only the effective stiffness
$k'$ associated with the second derivative of $U_w$ near its minimum 
is important.

In addition to reproducing the variation of $A_c$ with load,
this simple harmonic mean-field model makes several predictions about local forces
that are consistent with simulations.
One is that the distribution of instantaneous forces $f$ on
substrate atoms is exponential, which leads to a universal
distribution of the fraction of load born by the atoms with
the largest forces.
Figs.  \ref{ForceStaSnapshot} and \ref{exponentialforce}
show that these predictions describe contact
between flat surfaces with all geometries except at extremely
low temperatures ($\ll T_m/10$) and high pressures ($p/E^* >0.01$).

Another prediction is that for all atomic (commensurate, incommensurate,
amorphous, or stepped) and large scale (flat or sphere)
geometries, the mean force on an atom and the fraction of time in
contact have a functional relationship that depends only on temperature.
The same relation holds for adhesive surfaces if contact is defined
as repulsion and the mean repulsive force $f_c$ is calculated.
The fraction of time in contact rises linearly with force at low
forces and this quantitatively describes the variation of $A_c$
with load for commensurate surfaces.
For incommensurate and amorphous surfaces, the prediction need only be scaled
by a constant factor to include the fact that some substrate
atoms are too far from wall atoms to make contact.
The linear relation between $A_c$ and load or $p_c$ and $f_c$
breaks down when the fraction of time in contact approaches unity.
This does not occur until quite large values of $p/E^*$.
For example, $p_c$ reaches 50\% at $p/E^* = 0.01$ and $0.02$
for $T/T_m =0.1$ and $0.25$, respectively.

The simplicity of the harmonic mean-field model allows us to make estimates
about the range of validity of the above statements that are quite
general and should not depend on the details of atomic interactions.
Equation \ref{simple2} gives the ratio between
the dimensionless pressure and the fraction of area in contact
at any instant.
The main variation in this ratio comes from $\delta z_{\rm rms}$, which rises as
the square root of temperature.
The Lindemann criterion says that the value of $\delta z_{rms}$
should be about 10\% of the nearest-neighbor spacing at the
melting temperature \cite{footLind}.
This allows us to write
$\delta z_{rms} \approx 0.1 \sigma \sqrt{T/T_m}$.
Inserting this into Eq. \ref{simple2} and multiplying and dividing by
$k_{\rm eff}$ we find
\begin{equation}
\frac{A_c}{A_0} = c_A \frac{L}{A_0 E^*}= c_A \frac{p}{E^*}
\end{equation}
with
\begin{equation}
c_A = 10
\frac {E^* A_a} {\sigma k_{\rm eff}}
\frac{k_{\rm eff}}{k'} y 
\sqrt{\frac{T_m}{T}} \ \ .
\end{equation}
The values of $E^*$ and $k_{\rm eff}$ are both determined
by the interactions between neighbors, and the ratio
$E^* A_a/ \sigma k_{\rm eff}$ will generally be near unity.
For the potential used here it is $0.92$.
The value of $k'/k_{\rm eff}$ should also be close to unity if interactions
across the interface are comparable to internal interactions.
The value of $y$ is directly related to $A_c/A_0$ via
Eq. \ref{simple1}, but varies extremely
slowly.
For example, it drops
from 3.1 to 1.4 as $A_c/A_0$ rises from $10^{-3}$ to $10^{-1}$.
One can thus conclude that
\begin{equation}
c_A \sim  20 \sqrt{\frac{T_m}{T}}
\label{simple4}
\end{equation}
for any potential.
Note that this expression differs from the simple estimate
based on thermal collisions in Eq. \ref{csubA} by a factor of order unity.

Molecular dynamics is only accurate for temperatures high enough compared
to the Debye temperature that quantum effects can be ignored.
This typically corresponds to $T/T_m \gtrsim 0.05$ and applies to most materials
at room temperature.
As $T$ increases from $0.05 T_m$ to $T_m$ the pressure required for
atoms to be in contact more than 50\% of the time rises from $p/E^* \sim 0.01$
to 0.05.
Plastic deformation sets in when $p$ exceeds the hardness, $H$.
Typical values of $H/E^*$ are of order $10^{-4}$ to $10^{-2}$
for macroscopic crystals.
While larger values
may be observed in defect free systems and
at nanometer scales,
the theoretical limit is only $H/E^* \sim 0.1$.
Thus
one can generally expect that atoms will spend a significant
fraction of their time out of repulsive contact at
temperatures and pressures of experimental interest.

Our simulations reduced the number of free parameters and the size
of the simulation by keeping one surface rigid.
If both surfaces are compliant, there are two offsetting changes.
One is that there are thermal fluctuations by atoms on both sides
that increase thermal roughness.
This can be modeled roughly by replacing $k_{\rm eff}^{-1}$ by the
sum of the inverse values for the two surfaces.
Continuum contact mechanics says that the inverse of
the effective modulus is related in the same way to the moduli
of the two solids.
Since the same interactions determine both $E^*$ and $k_{\rm eff}$,
the two changes should nearly cancel.
Simulations for a few examples confirmed this.

The above discussion neglects quenched geometric disorder due to
differences in separation between substrate atoms and
atoms on the opposing surface.
This reduces the number of atoms
on amorphous and incommensurate surfaces
that can be brought into contact by thermal fluctuations.
At $T/T_m =0.1$ and 0.25 we found that the mean-field theory still worked
if one reduced the available number of contacts by a constant fraction.
For example, the fit lines in Fig. 5 used 84\% for the incommensurate
surface and 57\% for the amorphous case. 

At the much lower temperature of $10^{-4}\epsilon/k_{\rm B}$,
$\delta z_{\rm rms}\sim  0.001\sigma$ is much smaller than the atomic
scale roughness on incommensurate and amorphous surfaces.
Particularly in the case of amorphous surfaces, those atoms that
contact at a given load stay in contact most of the time.
This is clearly seen in Fig. 2, where
for all loads $A_c(\Delta t)$ increases by less than
a factor of 2 as $\Delta t/\tau$ increases from 0 to 500.
One can also calculate the mean force on atoms from the measured $A_c$ and $L$.
At the lowest load in Fig. 2c, $f_c \sim 0.2 \epsilon/\sigma$, which
is larger than the force $k' \delta z_{rms} \sim 0.1 \epsilon/\sigma$
where the linear relation between force and $p_c$ breaks down.
Thus, in contrast to the commensurate case, the rise in $A_c$
reflects an increase in the number of atoms that are close enough
to contact rather than an increase in the fraction of time
in contact.
As load increases,
contacting atoms are pushed down relative to other atoms,
allowing new atoms to contact the opposing surface.
Note that the rise in $A_c$ with load is nonlinear for
both incommensurate and amorphous surfaces.
Continuum theory predicts a linear relation for rough
surfaces, but it is not clear it can be applied at
these scales or that the surfaces have an appropriate
distribution of roughness.

Experimental surfaces are generally much rougher than those
considered here, with bumps on top of bumps at all scales \cite{greenwood66,bowden86}.
At low loads, contact occurs only where two asperities from
opposing surfaces overlap.
In continuum theories, the linear area-load relation comes
from the growth in the number of such contacts with load.
The distribution of contact sizes and forces remains unchanged.

Simulations of the spherical tip geometry give insight into
the behavior of asperity contacts.
As for flat surfaces, at typical experimental temperatures,
thermal roughness leads to a linear relation between
area and pressure until $p/E^*$ is 0.01 to 0.05.
This pressure is generally large enough to produce plastic deformation under
micrometer and larger asperities, and comparable to the
hardness of nanoasperities.
In the cases of nanoasperities, these and previous simulations
\cite{luan05,luan06,cheng10pre} show that the contact area
is not accurately described by continuum theory.
Thus it is not clear that $A_c$ will follow continuum
predictions \cite{bowden86,johnson85}
for rough surfaces at experimental temperatures.

Since continuum theory ignores thermal fluctuations, it is natural to
work with definitions of contact area based on time-averaged 
rather than instantaneous forces.
For adhesive surfaces, the area where time-averaged forces are repulsive
is nearly independent of averaging time and is insensitive to temperature.
While this seems the least ambiguous definition of contact, it only
shows a sharp transition from no contact to full contact for
flat commensurate surfaces.
For flat incommensurate and amorphous surfaces, the transition is spread
over a range $\Delta p /E^* \sim 0.03$ that is comparable to
the pressures needed to produce plasticity.

The time averaged force gives much less satisfactory results for
nonadhesive surfaces.
Since the force is always repulsive, any contact leads to a 
positive time-averaged force.
The contact area grows monotonically with the averaging time
interval $\Delta t$, leading to substantial ambiguity in $A_c$
for both flat and spherical surfaces.
Averaging positions instead of forces reduces the sensitivity
to rare events.
However, one finds that the mean separation is beyond the interaction
range for most loads, implying that there is no contact even though
there is a large repulsive force.

Averaging positions and forces gives similar results for adhesive surfaces
if contact is based on repulsive interactions between atoms.
However we have found that another definition gives a sharper transition
from partial to full contact for incommensurate surfaces.
Instead of basing contact on forces between atoms,
it is based on the separation between surfaces.
A Delaunay triangulation of each surface is performed and triangles
on opposing surfaces
are said to contact if they are separated by less than the distance
where atomic interactions become repulsive.
We are currently exploring whether associating contact with surface
separations instead of atomic forces 
improves the comparison of simulations and continuum calculations
of rough surfaces with multiasperity contacts.

\section*{Acknowledgments}

This material is based upon work supported by the National Science Foundation
under Grant No.~DMR-0454947 and the Air Force Office of Scientific
Research under Grant No.~FA9550-0910232.


\begin{thebibliography}{10}
\bibitem{footrepulse}
Indirect long-range replulsions may be present in some cases, such as for
  charged srufaces in water.
\bibitem{footrepel2}
Note that as in the theoretical expressions, only repulsive contributions to
  the force are included in the average.
\end{thebibliography}
\end{document}